# A Prototype of Reconfigurable Intelligent Surface with Continuous Control of the Reflection Phase
## Modeling, Full-Wave Electromagnetic Characterization, Experimental Validation, and Application to Ambient Backscatter Communications

R. Fara[1,2], P. Ratajczak[2], D.-T. Phan-Huy[2], A.Ourir[3], M. Di Renzo[1] and J. De Rosny[3]

[1]*Université Paris-Saclay, CNRS, CentraleSupélec, Laboratoire des Signaux et Systèmes,*
3 rue Joliot-Curie, 91192, Gif-sur-Yvette, France
[2]*Orange Labs Networks,* Châtillon & Sophia-Antipolis, France
[3]*ESPCI Paris, PSL University, CNRS, Institut Langevin*, Paris, France
romain.fara@orange.com

*Abstract*—With the development of the next generation of mobile networks, new research challenges have emerged, and new technologies have been proposed to face them. On the one hand, the reconfigurable intelligent surface (RIS) technology is being investigated for partially controlling the wireless channels. The RIS is a promising technology for improving the signal quality by controlling the scattering of the electromagnetic waves in a nearly passive manner. On the other hand, ambient backscatter communications (AmBC) is another promising technology that is tailored for addressing the energy efficiency requirements for the Internet of Things (IoT). This technique enables low-power communications by backscattering ambient signals and, thus, reusing existing electromagnetic waves for communications. RIS technology can be utilized in the context of AmBC for improving the system performance. In this paper, we report a prototype of an RIS that offers the capability of controlling the phase shift of the reflected waves in a continuous manner, and we characterize its characteristics by using full-wave simulations and through experimental measurements. Specifically, we introduce a phase shift model for predicting the signal reflected by the RIS prototype. We apply the proposed model for optimizing an RIS-assisted AmBC system and we demonstrate that the use of an RIS can significantly improve the system performance.

## Introduction

The development of each generation of mobile networks has improved the efficiency of wireless communications. From the second generation (2G) to the fifth generation (5G) of mobile networks, the performance of wireless communication systems has improved, thanks to research and development efforts on the design and optimization of the transmitters and receivers. However, the propagation channel separating the transmitters and receivers has inherently been perceived as not controllable and sometimes unfavorable to the propagation of the electromagnetic (EM) waves. This is one of the challenges that have been identified for next generation mobile networks [1].

Recently, the reconfigurable intelligent surface (RIS) technology has been proposed and investigated for application to wireless communications [2]. An RIS is a planar structure that comprises a large number of unit cells that are capable of scattering (e.g., reflecting) the EM signals in a controlled manner [3]. An RIS can modify the phase-shift, the amplitude or even the polarization of the incident signal thanks to a smart controller that tunes the response of each unit cell. Notably, the RIS technology is nearly passive, as it is completely based on the scattering of the EM waves and does not require power amplifiers for signal transmission. Indeed, some energy is only required for the smart controller and for enabling the reconfigurability of the RIS. Therefore, an RIS can mitigate the interference or improve the signal quality at some specific and localized network locations. The RIS technology shows promising potential for application in future networks, as it can partially control and shape the propagation channels as one desires. The deployment of RISs can, therefore, improve the signal reliability in a nearly passive manner, without additional densification of the network elements and without the use of active antennas at the transmitters and receivers. RISs can be combined with existing technologies like multiple-input multiple-output (MIMO) systems, millimeter-wave (mmWave) communications, terahertz (THz) communications, machine learning (ML) and artificial intelligence (AI) for enhancing the performance of sixth generation (6G) networks [3].

In particular, the RIS technology can be employed for enhancing the performance of another promising technology, especially in the context of the Internet of Things (IoT): ambient backscatter communications (AmBC) [4]. In AmBC, a device, named tag, transmits data to another device, named reader, without generating additional EM waves. In simple terms, the tag is illuminated by an ambient signal source, such as a TV tower, a WiFi hotspot, or even a base station, and such received signal is then backscattered by the tag. The tag is able to modulate the backscattered signal by implicitly transmitting its own message. The modulated signal is detected by the reader, which can decode the tag's message. A simple tag modulates the backscattered signal according to two possible states: (i) a backscattering state, when the tag antenna is short-circuited, and (ii) a transparent state, when the tag antenna is open-circuited

[5]. Since AmBC systems inherently reuse the ambient signals and do not transmit additional EM waves for communications, they are considered a suitable technology for very low power applications [6]. However, the tag-to-reader link may be inherently interfered by the ambient signal. Also, the tag may not backscatter a strong enough signal to the reader, if, e.g., it is located in a deep fade of the ambient signal [7]. AmBC is, therefore, a promising technology, yet it has some limitations.

To overcome the just mentioned limitations, AmBC can be assisted by an RIS. For example, the authors of [8] show that the phase shifts of an RIS can be optimized to compensate for the multipath effect of the propagation channel, in order to reduce the interference and to increase the signal quality at the reader. In [9], an RIS plays the role of the tag and is used to backscatter the message to the reader. Since many unit cells are available at the RIS, the backscattered signal is enhanced and can be more easily detected. This principle is demonstrated experimentally in [10], and it is shown to significantly improve the system performance. Also, the authors of [11] report that, in cognitive networks, an RIS can improve the performance of the tag-to-reader link by maximizing the signal at the tag's location [12].

Therefore, recent research works have demonstrated that the RIS technology is a suitable candidate for improving the performance of AmBC. In [12], in particular, we have experimentally demonstrated that the tag-to-reader link can be enhanced by creating a passive reflected beam that improves the signal quality at the tag's and reader's locations. Similar to MIMO systems [13] but in a nearly passive manner, an RIS beamforms the reflected signal to assist the transmission of the tag. The experiments conducted in [12] utilize an RIS prototype that comprises 196 reconfigurable unit cells, each capable of phase-shifting the reflecting signal in a controlled manner. This is realized by using varactor diodes that control the phase response of the unit cells, which enables a continuous tuning of the phase shifts of the reflected signal. Even though the principle has been experimentally validated in [12], the RIS has not been completely characterized and, in particular, no equivalent model for the phase shift and amplitude response of the reflection coefficient was given. This is, however, essential for utilizing the RIS prototype in wireless communications.

In this paper, we focus on this specific and important aspect, and introduce a simplified model for the phase and amplitude response of the reflection coefficient of a voltage controlled RIS prototype. Based on the proposed model, we evaluate its suitability and accuracy in the context of RIS-assisted AmBC. In particular, the validation of the model is performed through full-wave simulations in Ansys HFSS and through experimental measurements. We demonstrate that the complex EM behavior of each unit cell can be modeled with simple equations that depend on the control voltage applied to the unit cells. Based on the proposed model for the unit cell, we propose a method to create a large codebook to realize passive beamforming. For given locations of the source, tag, and reader, the unit cells of the RIS are appropriately tuned to reflect and maximize the signal towards desired spots. We demonstrate that the signal scattered by the RIS can be well approximated by the linear combination of the contribution of each unit cell (based on the proposed model). Also, we prove that an RIS with a continuous phase-shift tuning capability can improve the performance of AmBC. We discuss the accuracy of the proposed (simple) model for the phase shift and compare it, with the aid of full-wave simulations, against more complex models that account for a finer-grained modeling of the response of the unit cells. We show that the proposed simplified model is sufficiently accurate for first-order analysis of RIS-assisted communications within a given range for the angles of incidence and reflection.

The paper is organized as follows. First, we introduce the RIS prototype with continuous phase-shifting capability and discuss its characteristics. Then, we describe a method to define a large codebook for realizing passive beamforming through controllable phase-shifting. Subsequently, we illustrate the RIS-assisted AmBC principle and compare the predictions from the proposed phase shift model against accurate full-wave simulations with Ansys HFSS. Finally, we conclude the paper.

## RIS Prototype with Continuous Phase-Shifting

As illustrated in Fig. 1, the RIS prototype comprises 14×14 unit cells. Each unit cell is a reflecting conductive patch that can be reconfigured independently of the others. In particular, the unit cell is a three-dimensional patch on a dielectric substrate. The patch is made of two conductive areas separated by an annular slot. The inner and outer areas are connected to each other with four MA46H120 varactor diodes. The specific arrangement of the varactor diodes ensures the adjustment of the capacitance of the unit cell, which provides the desired tunability of the reflection coefficient of the unit cell. In detail, in the reverse configuration, the capacitance of a varactor diode is inversely proportional to the junction voltage. Thus, the capacitance of the unit cell can be tuned by controlling the voltage applied to the four varactors. As illustrated in Fig. 1, the unit cell can be reconfigured by applying different voltages to the four varactor diodes, which, in turn, enables the configuration of the reflection coefficient.

In the realized unit cell structure, four varactors are used for each unit cell for ensuring a good symmetry of the radiation pattern. All unit cells have identical design and size, i.e., 14×14 mm. They are engineered to operate in the frequency range 5.15-5.75 GHz [13, 14], which yields a cell size of approximately a quarter of the signal wavelength. The unit cell structure is characterized through full-wave simulations in Ansys HFSS under periodic boundary conditions. In detail, the results are obtained by considering an infinitely large surface that is a periodic repetition of a unit cell configured to realize a given reflection coefficient. Thus, a homogeneous infinite surface is simulated with the desired unit cell configuration, which ensures the correct definition of reflection coefficient.

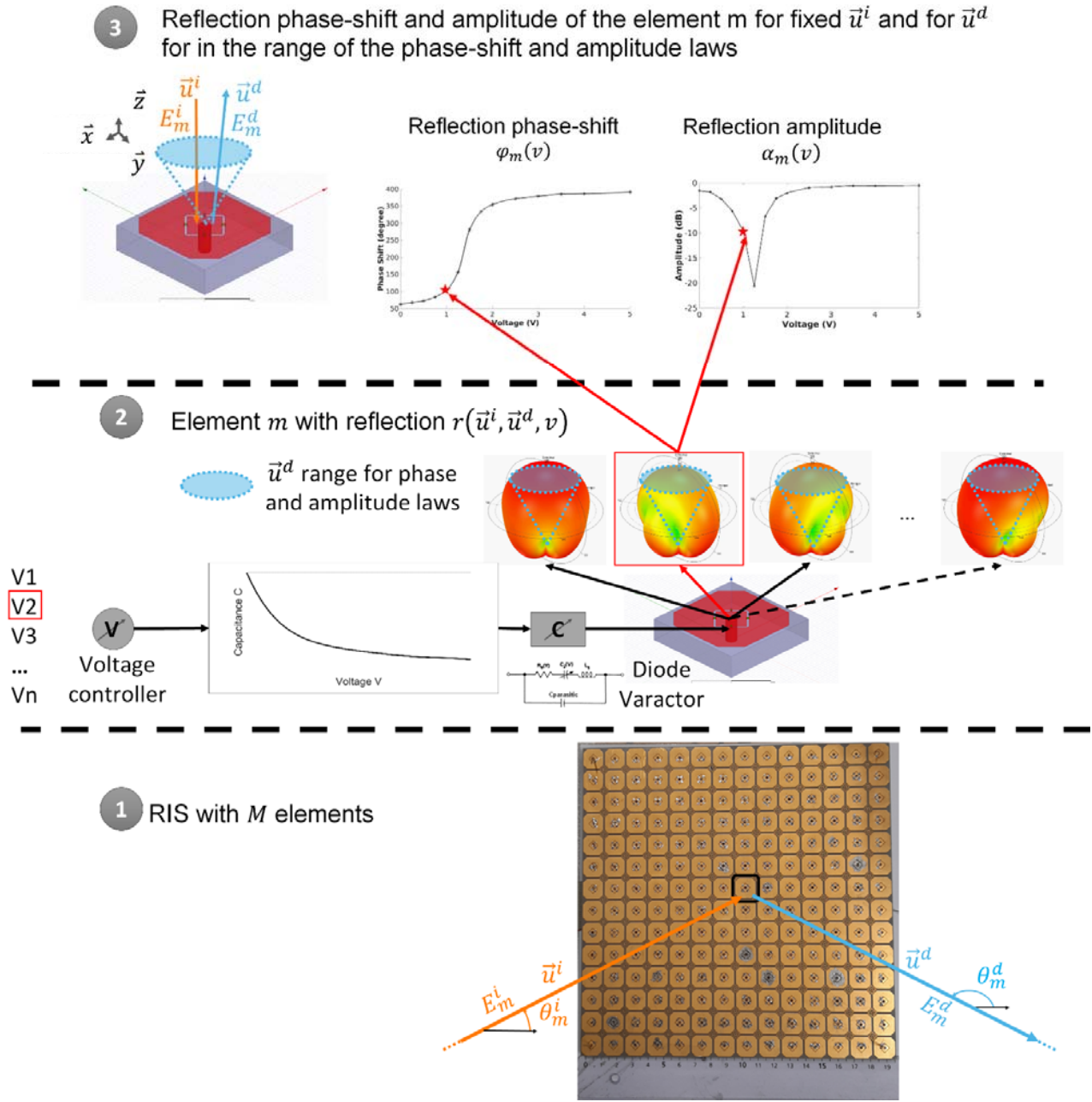

Fig. 1. Unit cell with a voltage-controlled tunable reflection coefficient $r(\vec{u}^i, \vec{u}^d, v)$. Full-wave simulation results under periodic boundary conditions [14, 15]

As illustrated in Fig. 1, the desired reflection coefficient of each unit cell can be obtained through the voltage-controlled capacitance that characterizes each unit cell. More precisely, if the $m^{th}$ unit cell is illuminated by an electric field $E^i$ whose direction of incidence is $\vec{u}^i$, the reflected electric field towards the direction of departure $\vec{u}^d$ can be written as $E^d = E^i \cdot r(\vec{u}^i, \vec{u}^d, v)$, where $v$ is the voltage applied to the unit cell, $\vec{u}^i$ and $\vec{u}^d$ are unit-norm vectors defined by the azimuth and elevation angles of incidence $(\theta^i, \varphi^i)$ and departure $(\theta^d, \varphi^d)$, respectively, in spherical coordinates, and $r(\cdot)$ is the reflection coefficient.

To characterize the unit cell and to introduce an EM-consistent model for $r(\vec{u}^i, \vec{u}^d, v)$ that can be utilized in wireless communications, we study the scattering properties of the unit cell, as a function of $\vec{u}^i, \vec{u}^d$, and $v$, by using the Ansys HFSS EM software under periodic boundary conditions and through measurements conducted with the RIS prototype. As far as the full-wave simulations are concerned, the reflection coefficient of each unit cell is indirectly characterized as a function of the capacitance $c$ of the varactor diodes. As far as the experimental characterization through measurements is concerned, the reflection coefficient of each unit cell is directly characterized as a function of the voltage $v$ applied to the varactor diodes. Since the capacity of each varactor diode is voltage-controlled, as illustrated in Fig. 1, $c$ and $v$ are directly related to each other, i.e., $v = v(c)$ or $c = c(v)$. This relation is not necessary in this paper, but it can be derived through measurements.

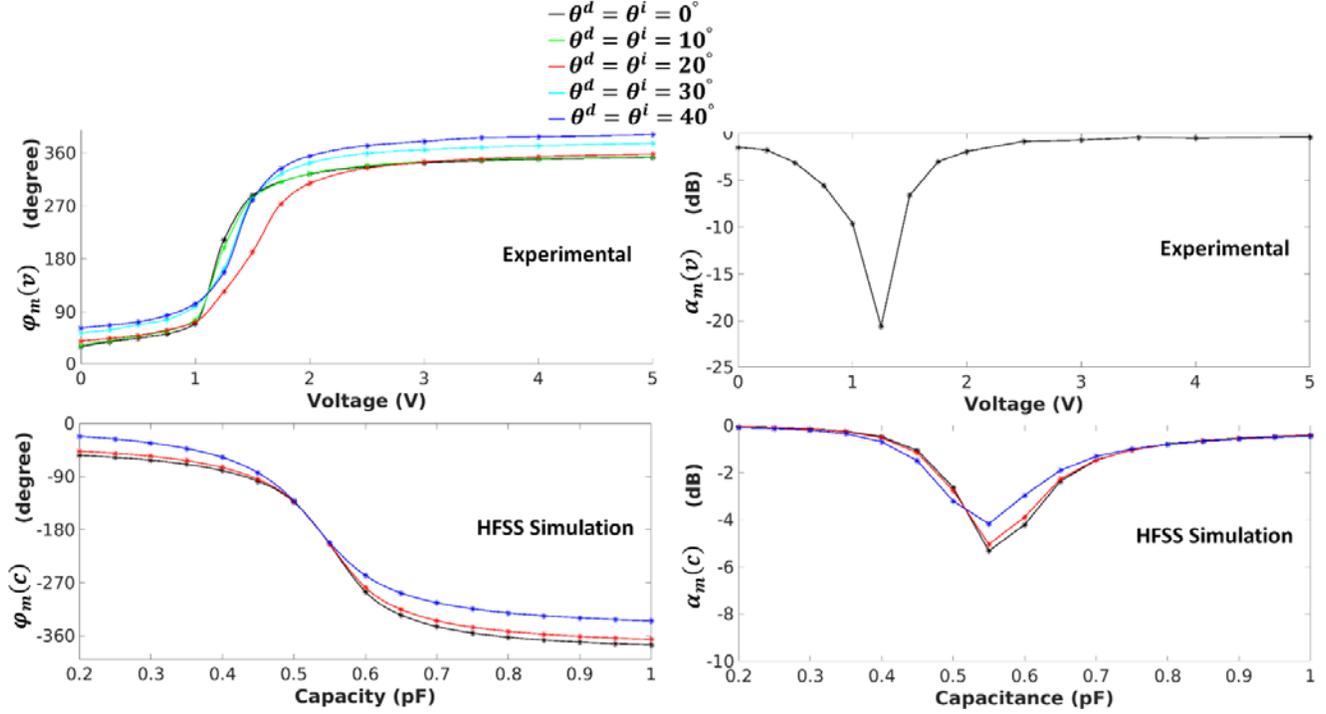

Fig. 2. Phase and amplitude of $r(\vec{u}^i, \vec{u}^d, v)$ as a function of $v$ and for fixed $\vec{u}^i$ and $\vec{u}^d$ ($\varphi^i = 0$, $\varphi^d = 180$ degrees). HFSS simulations vs. experimetal measurements.

The characterization of the reflection characteristics of the unit cell is illustrated in Fig. 2 for various elevation angles of the incident and reflected electric field. The figure reports, in particular, the phase $\varphi_m(\cdot)$ and the amplitude $\alpha_m(\cdot)$ of the reflection coefficient as a function of the capacitance of the varactor diodes (HFSS simulations) and the junction voltage (prototype experiments). From the results illustrated in Fig. 2, we evince that both the phase and the amplitude of the reflection coefficient of each unit cell are almost independent of the angle of incidence and reflection, if the deflection angle is less than 40 degrees. Within this range of angles, notably, the radiation pattern of the unit cell can be assumed to be almost omnidirectional. For wider angles of deflection, on the other hand, Fig. 1 shows that the radiation pattern of the unit cell is not omnidirectional, but it depends on the angle of incidence and reflection. As expected, Fig. 2 shows that the reflection coefficient can be adjusted by appropriately controlling the junction voltage in the range 0-5 V, which corresponds to a capacity in the range 0.2-1 pF.

By considering, therefore, angles of incidence and reflection no larger than 40 degrees with respect to the normal to the unit cell, a simple model for the reflection coefficient of the unit cells can formulated as follows:

$$r(\vec{u}^i, \vec{u}^d, v) \approx r(v) \approx g_0 \alpha_m(v) e^{j\varphi_m(v)} \quad (1)$$

where $g_0$ denotes the unit cell (constant) gain, and $\alpha(v)$ and $\varphi(v)$ denote the amplitude and the phase of the unit cell reflection coefficient, respectively, which can be configured in a continuous manner as a function of the voltage $v$ that is applied to the varactor diodes.

The relation in (1) has been experimentally characterized with the aid of the RIS prototype and some experimental measurements. The obtained functions, $\alpha(v)$ and $\varphi(v)$, are reported in Table I as a function of the voltage $v$. Table I is valid only for angles less than 40 degrees for which the RIS was designed, optimized, and engineered.

TABLE I. EXPERIMENTAL CHARACTERIZATION OF THE REFLECTION COEFFICIENT IN (1): $\alpha(v)$ AND $\varphi(v)$ AS A FUNCTION OF $v$.

| Voltage [V] | $\alpha(v)$ [dB] | $\varphi(v)$ [degrees] |
|---|---|---|
| 0 | -1.517 | 32.798 |
| 0.25 | -1.807 | 40.854 |
| 0.5 | -3.156 | 46.807 |
| 0.75 | -5.59 | 53.543 |
| 1 | -9.576 | 70.32 |
| 1.25 | -20.563 | -167.158 |
| 1.5 | -6.615 | -73.171 |
| 1.75 | -3.029 | -49.627 |
| 2 | -1.959 | -35.908 |
| 2.5 | -0.874 | -23.263 |
| 3 | -0.749 | -16.087 |
| 3.5 | -0.469 | -12.663 |
| 4 | -0.528 | -9.925 |
| 5 | -0.439 | -6.906 |

By analyzing Table I, we observe that the amplitude and the phase of the reflection coefficient are not independent of each other, since they both depend on the control voltage. In particular, the amplitude of the reflection coefficient has the smallest value in correspondence of the range of major variation of the phase of the reflection coefficient (see Fig. 2).

In this section, we have characterized the phase and amplitude response of the RIS prototype, and have introduced a simple model for the reflection coefficient of each unit cell, based on full-wave simulations, under the modeling assumption of periodic boundary conditions, and experiments. We have demonstrated the main capabilities of the designed RIS prototype: The phase-shift induced by each unit cell can be continuously controlled as a function of the control voltage applied to the unit cell. This feature can be exploited in order to control the angle of reflection of the incidence signal and to beamform the reflected signal towards the desired direction. This is analyzed and experimentally validated next.

**RIS CODEBOOK DESIGN FOR ACCURATE BEAMSTEERING**

From the EM characterization of the unit cell in Fig. 1 and Fig. 2, we evince that it is possible to treat an RIS as a black-box that comprises several identical and voltage-controlled unit cells, each capable of independently changing the amplitude and phase of the incident signal. By appropriately optimizing all the unit cells of the RIS, the incident signal can be partially controlled and steered towards desired locations, in order to enhance or null the received signal in a nearly passive manner, without power amplification, without decoding and reencoding the incident signal, and without requiring radio frequency chains. If the operation of the RIS prototype is restricted to angles of incidence and reflection smaller than 40 degrees (for which our prototype was designed), Table I constitutes the proxy between the EM world and the wireless communications world.

An appropriately optimized RIS can, therefore, steers the reflected signal towards desired directions or specified locations. Assuming, for example, that the locations of the transmitter and RIS are known, we can pre-design a codebook of beamforming vectors that provide the exact configuration of the phase shifts of every unit cell of the RIS in order to steer the reflected signal towards a receiver that is located in a given position. This is obtained by ensuring that the signals scattered by each unit cells add up coherently in correspondence of the desired receiver location. Then, given an area of interest, a location-dependent codebook for nearly passive beamforming can be engineered.

A simple approach to realize an RIS that focuses the reflected signal towards a specified location, in line-of-sight of both the RIS and the transmitter, is to ensure that (i) the phase shift applied by each unit cell is optimized to guarantee that the signals reflected by all the unit cells add up coherently at the location of the receiver and (ii) each unit cell adds an additional phase shift that ensures that signal reflected form the RIS adds up coherently with the signal emitted by the transmitter at the location of the receiver. Our proposed RIS prototype enables these two functionalities simultaneously. Based on Table I, this can be realized by controlling the junction voltage of each unit cell, so as to apply a cell specific phase shift and an addition constant RIS phase shift.

More precisely, let $E^S$ be the transmitted electric field at the transmitter (source) location $S$ and let $E^P$ be the RIS-reflected electric field at the predefined location $P$. Based on (1), under the assumption of free-space propagation, we have:

$$E^P = E^S \sum_{m=1}^{M} \frac{\lambda e^{j\left(2\pi \frac{d_m^{S-RIS}}{\lambda}\right)}}{4\pi d_m^{S-RIS}} [g_0 \alpha_m(v) e^{j\varphi_m(v)}] \frac{\lambda e^{j\left(2\pi \frac{d_m^{Ris-P}}{\lambda}\right)}}{4\pi d_m^{Ris-P}} \quad (2)$$

where $M$ is the total number of unit cells of the RIS, $d_m^{S-RIS}$ and $d_m^{RIS-P}$ are the transmission distances between the source and the $m^{th}$ cell of the RIS, and between the $m^{th}$ cell of the RIS and the predetermined target location $P$, respectively, and $\lambda$ is the signal wavelength. Finally, $j$ denotes the imaginary unit.

From (2), we evince that the phase shift that corresponds to the optical path-length associated with the $m^{th}$ unit cell at the target location $P$ is as follows:

$$b_m^{(P)} = e^{-j2\pi \frac{d_m^{S-Ris} + d_m^{Ris-P}}{\lambda}} \quad (3)$$

From (3), in order to ensure that the signals scattered by all the unit cells are phase-aligned with the transmitted signal at the target location $P$, the phase shift that each unit cell of the RIS needs to apply to the incidence signal is $\varphi_m = b_m^{(P)} - \psi^{(P)}$, where $\psi^{(P)}$ is the uniform phase of the transmitted signal observed at the target location $P$.

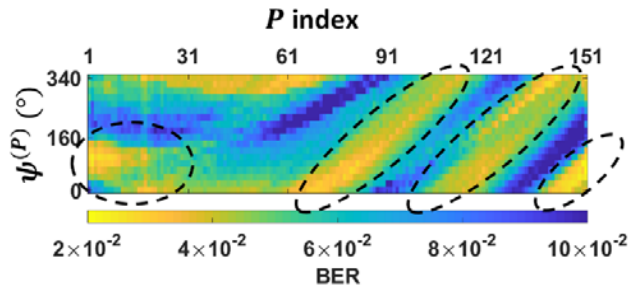

Fig. 3. Experimental measurements of the BER for all possble values of the predefined locations ($p=1,2,...,151$) and phase shifts $\psi^{(P)}$ (in the range 0-360 degrees). The experimetnal setup is illustrated in Fig. 4.

An illustration of the bit error rate (BER) performance of an RIS-assisted communication system, deployed by using our RIS prototype, is illustrated in Fig. 3 as a function of several target locations ($P$ index) and RIS phase shifts $\psi^{(P)}$. A photo of the considered system setup for application to AmBC is given in Fig. 4. The experimental results in Fig. 3 are obtained by pre-designing the beamforming codebook in (3) in correspondence of 151 possible locations of a target receiver, and by then measuring the BER, at a specified location, by testing the 151 pairs of phase shifts $(b_m^{(P)}, \psi^{(P)})$ of the RIS codebook. From Fig. 3, we observe that there are multiple pairs $(b_m^{(P)}, \psi^{(P)})$ that provide good BER performance. In particular, the pairs depicted in yellow correspond to possible configurations of the RIS that enchance the link performance, as compared to the same system setup in the absence of RIS.

It is worth mentioning that the performance enhancement offered by RIS-assisted communications can be further improved (with respect to the results illustrated in Fig. 3), since the beamforming design in (3) is sub-optimal and the simplest one that can be applied, since it accounts only for the different optical path-lengths of the signal scattered by the unit cells.

**RIS-ASSISTED AMBC: UNIT CELL MODELING ACCURACY**

In AmBC, the system performance highly depends on the power level of the signal received at the tag. Since a tag is a passive device, the amount of power that is backscattered is ultimately determined by the amount of power that the tag receives from the transmitter (the source in Fig. 4). In this context, an RIS can be utilized to passively beamform the signal emitted by the source towards the tag, so as to enhance the amount of power received and, therefore, the amount of power that is backscattered towards the receiver (the reader in Fig. 4). To this end, the RIS needs to be appropriately optimized in terms of beamforming vector and uniform phase shift, i.e., the optimal pair $(b_m^{(P)}, \psi^{(P)})$ needs to be identified, where $P$ denotes the location of the tag in the considered example.

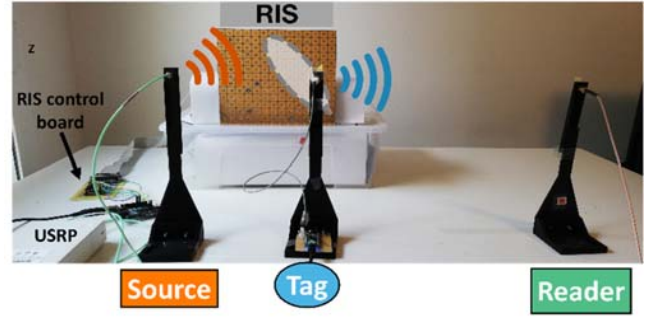

Fig. 4. Illustration of the considered testbed for RIS-assisted AmBC.

Therefore, it is important to use an accurate model for the RIS in order to enable a simple but accurate design of the configuration of the unit cells. In Table I, we have introduced a simple model for the amplitude and phase of the reflection coefficient of each individual unit cell of the RIS, as a function of the control voltage. The objective of this section is to evaluate the accuracy of the model introduced in Table I, as compared with full-wave simulations performed with Ansys HFSS. The results of this analysis are illustrated in Fig. 5.

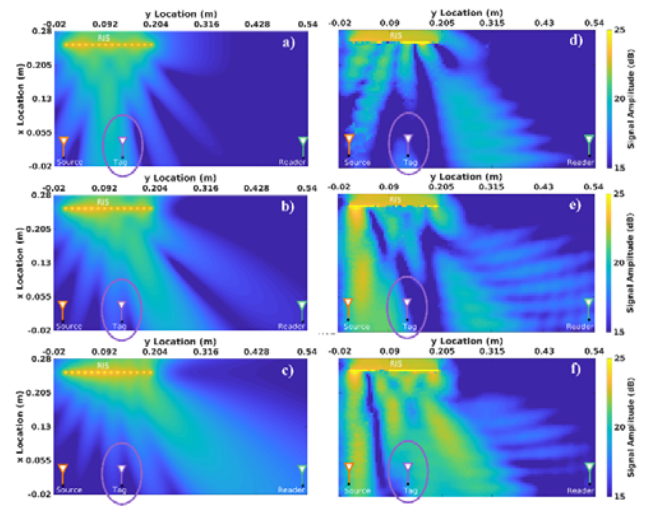

Fig. 5. Illustration of the field scattered by the RIS for three different beamforming pairs $(b_m^{(P)}, \psi^{(P)})$ that mininize the BER in Fig. 3: (a)-(c) shows the results obtained with the simple model in Table I and (d)-(f) show the results obtained through full-wave simulations in Ansys HFSS.

The results reported in Fig. 5 are obtained as follows.

- *Simplified model in Table I*. Given the locations of the source, tag, RIS, and reader in Fig. 4, three pairs $(b_m^{(tag)}, \psi^{(tag)})$ that minimize the BER in Fig. 3 are considered. The optimal beamforming vector and uniform phase shift are identified by using the simplified model for the RIS in Table I, and the simple analytical model for the scattered field in (2). In particular, we assume $g_0 = 1$ but we account for the fact that the amplitude of the reflection coefficient of each unit cell is not independent of the applied phase shift (and the unit cell control voltage).

- **Full-wave simulations in Ansys HFSS.** We simulate the entire RIS structure in Ansys HFSS by considering the actual scattering from each unit cell as illustrated in Fig. 1. The same three pairs $\left(b_m^{(tag)}, \psi^{(tag)}\right)$ as for the simplified model are considered, but no assumption on the modeling of the unit cells is made. In particular, the analytical models in Table I and (2) are not assumed, and the mutual coupling among the unit cells of the RIS is considered. In addition, the finite size of the RIS is taken into account in the full-wave simulations and no periodic boundary conditions at the unit cell level are applied in this case.

The results illustrated in Fig. 5 show, as expected, that some differences exist between the predictions obtained by using Table I and (3), as compared with those obtained with accurate full-wave simulations. We note, however, that the simplified model for the RIS in Table I well predicts the direction of the reflected beam of interest. More precisely, we observe that a hot spot can be clearly observed in correspondence of the location of the tag (circled in purpled color in Fig. 5), which confirms the signal enhancement provided by the RIS at the target location. Also, this is in agreement with the experimental results reported in Fig. 3. The mismatch between the simple model in Table I and the full-wave simulations in Ansys HFSS can be reduced by enriching the model for the RIS proposed in Table I at the expenses of a higher modeling and optimization complexity. Also, it needs to be mentioned that the tag in Fig. 5 is located near the boundary of Fraunhofer's far-field, which in part justifies the mismatch in Fig. 5.

## CONCLUSION

In this paper, we have described a prototype of an RIS whose unit cells that can be digitally and individually tuned through a control voltage. By appropriately configuring the voltage, the phase of the scattered field of each unit cell can be optimized. We have characterized the electromagnetic behavior of the individual unit cells of the RIS under the assumption of periodic boundary conditions, and have proposed a simple voltage-dependent model for the reflection coefficient of the unit cells. If the angles of incidence and reflection are confined in the range ±40 degrees, we have shown that the unit cells can be assumed to have an almost omnidirectional and unit-gain radiation pattern, while the amplitude of the reflection coefficient depends on the applied phase shift through the control voltage. We have compared the proposed simple model against full-wave simulations and have shown that, despite simple, the proposed model can be successfully employed for first-order optimization of the RIS. Finally, by capitalizing on the proposed model for the RIS and method for optimizing its nearly passive beamforming codebook, we have substantiated through experimental measurements that an RIS can enhance the performance and reliability of AmBC systems.

## ACKNOWLEDGMENT

This work has been supported in part by the European Commission through the H2020 RISE-6G project under grant agreement number 101017011.